\documentclass[prb,twocolumn,amsmath,amssymb,floatfix,showpacs,superscriptaddress]{revtex4}

\usepackage{graphicx}
\usepackage[]{epsfig}
\begin{document}

\draft

\def\sigmav{{\mbox{\boldmath{$\sigma$}}}}

\title{Spin filtering due to quantum interference in periodic mesoscopic networks}

\author{Amnon Aharony }
\address{Department of Physics and the Ilse Katz Center for
Meso- and Nano-Scale Science and Technology, \\Ben Gurion
University, Beer Sheva 84105, Israel} \author{Ora Entin-Wohlman }
\address{Department of Physics and the Ilse Katz Center for
Meso- and Nano-Scale Science and Technology, \\Ben Gurion
University, Beer Sheva 84105, Israel}
\author{Yasuhiro Tokura}
\address{NTT Basic Research Laboratories,
NTT Corporation, Atsugi-shi, Kanagawa 243-0198, Japan}
\author{ and Shingo Katsumoto}
\address{Institute of Solid State Physics, University of Tokyo, Kashiwa, Chiba 277-8581, Japan}

\begin{abstract}
We present several new results, extending our recent proposal of a
spin filter based on a tight-binding model for a periodic chain of
diamond-like loops [Phys. Rev. B {\bf 78}, 125328 (2008)]. In this
filter, the Rashba spin-orbit interaction (which can be tuned by a
perpendicular gate voltage) and the Aharonov-Bohm flux (due to a
perpendicular magnetic field) combine to select only one
propagating ballistic mode. For this mode, the electronic spins
are fully polarized along a direction that can be controlled by
the electric and magnetic fields and by the electron energy. All
the other modes are evanescent. Generalizing the square diamonds
into rhombi with arbitrary opening angles, we find that increasing
these angles widens the parameter range for efficient filtering. A
different gate voltage on the two sides of each rhombus is found
to delocalize the electrons for energies on one side of the band
center. We also compare our tight-binding model with models which
use continuous quantum networks of one-dimensional wires, and find
coincidence only when one chooses particular site energies at the
nodes of the network.
\end{abstract}

\date{\today}
\maketitle


\section{Introduction}

 Future device technology and quantum information
processing may be based on spintronics \cite{1}, where one
manipulates the electron's spin (and not only its charge).   Here
we address attempts to build mesoscopic spin filters (or spin
valves), which generate a tunable spin-polarized current out of
unpolarized electron sources. Much recent effort in this direction
uses narrow-gap semiconductor heterostructures, where the spins
are subject to the Rashba \cite{3} spin-orbit interaction (SOI):
in a two-dimensional electron gas confined by an asymmetric
potential well, the strength of this SOI can be varied by an
electric field perpendicular to the plane in which the electrons
move \cite{koga}. An early proposal of a spin field-effect
transistor \cite{2} used the Rashba SOI to control the spin
precession of electrons moving in quasi-one-dimensional wires.

Some of the most striking quantum effects arise due to
interference, which is best demonstrated in quantum networks
containing loops. Indeed, interference due to the Rashba SOI has
been measured on a nanolithographically-defined square loop array
\cite{koga06}. Recently, several theoretical groups proposed spin
filters based on a {\it single} loop, subject to both an electric
and a magnetic (Aharonov-Bohm (AB) \cite{AB}) perpendicular fields
(e.g Refs. \cite{citro,molnar,hatano,oreg}). However, such devices
produce a full polarization of the outgoing electrons only for
{\it special values} of the two fields. Later work considered the
effects of the Rashba SOI on the conductance of chains of loops.
These included studies of chains of diamond-like loops
\cite{berc1,berc2}, and studies of finite chains of circular loops
\cite{molnar2}. Although both studies showed some destructive
interference due to the SOI, they did not concentrate on the
tuning of the fully polarized spins which can be obtained in
certain parameter ranges.

\begin{figure}[ht]
\vspace{-.4cm}
\begin{center}
\includegraphics[width=6 cm]{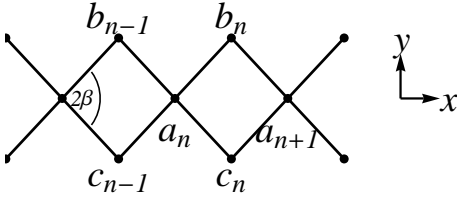}
\vspace{-.6cm}
\end{center} \caption{Chain of diamonds.}\label{1}
\end{figure}

Recently \cite{PRB}, we proposed a spin filter based on a periodic
chain of diamond-like square loops, connected to each other at
opposite corners [see Fig. \ref{1}]. Unlike the above earlier
papers, which used a continuum description for the wires on the
network, we used a simple tight-binding model, with quantum dots
(or `atoms') only at the nodes of the square diamonds. This
allowed us to obtain transparent analytical expressions for the
ballistic conductance through the chain and for the outgoing spin
polarization. We found that a combination of both the Rashba SOI
and the AB flux through each loop can result in destructive
interference, which can block the transmission of all the spin
components except one, which is then polarized at a tunable
direction. Technically, this single spin direction is associated
with a single propagating wave solution of the Schr\"odinger
equation, while all the other solutions involve evanescent modes.

 Here we extend our analysis of this
diamond-like chain in several directions. First, we replace the
previous square loops by rhombi, with a general angle $2\beta$
(Fig. \ref{1}). It turns out that the filter is more efficient for
$\beta>\pi/4$. Second, we generalize our previous study, by
allowing different site energies (controlled by appropriate gate
voltages) on the various sites in the unit cell (i.e. sites $a,\
b$ and $c$ in Fig. \ref{1}). Different site energies at sites $b$
and $c$ turn out to have drastic effects on the ballistic
conductance. Third, we propose using this filter at fixed electric
and magnetic fields, controlling the outgoing polarization using a
gate voltage. (In Ref. \cite{PRB} we worked at fixed energy, and
varied the magnetic and electric fields.) Fourth, we replace each
edge of each rhombus by a tight-binding chain of `atoms' (or
quantum dots). In this context, we compare our tight-binding
approach with the continuous quantum network approach used in
earlier work on the same geometry \cite{berc1,berc2}. As we
discuss elsewhere \cite{JPC}, the two approaches are not
equivalent.

Section 2 outlines the tight-binding model which we use for
solving the Schr\"odinger equation on the periodic chain of
generalized diamonds. Section 3 presents results for the
 polarization of the electrons in
the regions where they are fully polarized. We compare our
tight-binding approach to earlier continuum network models in Sec.
4, and summarize our results in Sec. 5.

\section{Generalized tight-binding model}

The basic theory was presented in Ref. \cite{PRB}. However, for
the generalizations introduced here we find it easier to use the
(rotated) coordinate axes shown in Fig. \ref{1}. Setting the
opening angle of each diamond to $2\beta$, the coordinates of the
sites in the $n$'th unit cell become ${\bf r}(a_n)=(n{\bar L},\
0,\ 0),\ {\bf r}(b_n)={\bf r}(a_n)+(L\cos\beta,L\sin\beta,0)$ and
${\bf r}(c_n)={\bf r}(a_n)+(L\cos\beta,-L\sin\beta,0)$, where
${\bar L}=2L\cos(\beta)$ is the basic step along the $x-$axis. We
start with the simplest tight-binding model, which has `atomic'
sites only at the corners of the diamonds (this will be extended
below). The hopping $2 \times 2$ unitary matrices within the
$n$'th diamond thus become
\begin{align} U^{}_{ab}(n)&=e^{in\phi/2+i\alpha\sigma^{}_1},\
\ \ U^{}_{ac}(n)=e^{-in\phi/2-i\alpha\sigma^{}_2},\nonumber\\
U^{}_{ba'}(n)&=e^{-i(n+1)\phi/2-i\alpha\sigma^{}_2},\nonumber\\
U^{}_{ca'}(n)&=e^{i(n+1)\phi/2+i\alpha\sigma^{}_1},\label{eq1}
\end{align}
where $a'$ denotes the site $a_{n+1}$, ${\bf \sigmav}$ is the
vector of Pauli matrices, $\alpha=k^{}_{SO}L$ ($k^{}_{SO}$
represents the strength of the Rashba term),
$\sigma^{}_1=\sigma^{}_x\sin\beta-\sigma^{}_y\cos\beta$,
$\sigma^{}_2=\sigma^{}_x\sin\beta+\sigma^{}_y\cos\beta$ and
$\phi/(2\pi)= B L^2 \sin(2\beta)/\Phi_0$ is the number of flux
units through each diamond. For each bond, the hopping matrix must
be multiplied by the hopping energy, $J_{uv}=J$ (below we measure
all energies in units of $J$).

Within the $n$'th diamond, the Schr\"odinger equations for the
spinors $\psi_a(n),\ \psi_b(n)$ and $\psi_c(n)$ are
\begin{align}
&(\epsilon-\epsilon^{}_a)\psi^{}_a(n)=-J\bigl
(U^{}_{ab}(n)\psi_b(n)+U^{}_{ac}(n)\psi_c(n)\nonumber\\
&+U^{\dagger}_{ba'}(n-1)\psi_b(n-1)+U^{\dagger}_{ca'}(n-1)\psi_c(n-1)
\bigr
),\nonumber\\
 &(\epsilon-\epsilon^{}_b)\psi_b(n)=-J\bigl
(U^\dagger_{ab}(n)\psi_a(n)+U^{}_{ba'}(n)\psi_a(n+1)\bigr ),\nonumber\\
&(\epsilon-\epsilon^{}_c)\psi_c(n)=-J\bigl
(U^\dagger_{ac}(n)\psi_a(n)+U^{}_{ca'}(n)\psi_a(n+1)\bigr
).\label{eq2}
\end{align}
Except for the special energies $\epsilon=\epsilon^{}_b,\
\epsilon^{}_c$, which represent dispersionless solutions (not
shown in the figures), we express $\psi_b(n)$ and $\psi_c(n)$ in
terms of $\psi_a(n)$ and $\psi_a(n+1)$, and substitute into the
equations for $\psi_a(n)$. We end up with effective
one-dimensional equations,
\begin{align}
4\lambda\Psi_a(n)={\bf W}^\dagger\Psi_a(n-1)+{\bf
W}\Psi_a(n+1),\label{ren}
\end{align}
with $4\lambda=\epsilon-\epsilon^{}_a-2\gamma^{}_b-2\gamma^{}_c$,
 $\gamma^{}_j=J^2/(\epsilon-\epsilon_j),\ j=b,c$, and with the
 non-unitary $2\times 2$ matrix
\begin{align}
&{\bf W}=\gamma^{}_bU^{}_{ab}(n)U^{}_{ba'}(n)+\gamma^{}_cU^{}_{ac}(n)U^{}_{ca'}(n)\nonumber\\
&\equiv 2(d-ib_y\sigma^{}_y-b_z\sigma^{}_z);\ \ \
2d=a^{}_+[c^2-s^2\cos(2\beta)],\nonumber\\
&2b_y=2a^{}_+ c s \cos\beta,\ \ \
2b_z=-ia^{}_-s^2\sin(2\beta),\label{db}
\end{align}
with $c=\cos\alpha$, $s=\sin\alpha$ and 
$a^{}_\pm=\gamma^{}_be^{-i\phi/2}\pm\gamma^{}_ce^{i\phi/2}.$

\begin{figure}[ht]
\begin{center}
\includegraphics[width=6cm]{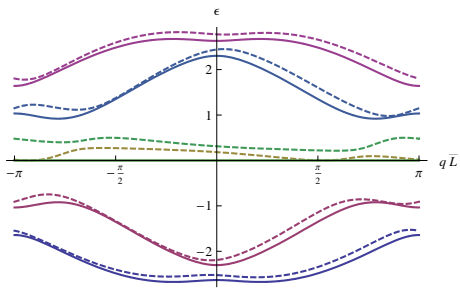}
\end{center} \caption{Spectra for $\beta=\pi/4,\ \phi=.4\pi,
\ \alpha=.2\pi$. Four full lines: $\epsilon^{}_b=\epsilon^{}_c=0$.
Six dashed lines: $\epsilon^{}_b=.5,\ \epsilon^{}_c=0$. All
energies are in units of $J$.}\label{spec}
\end{figure}

Assuming a propagating wave, $\Psi_a(n)=Ce^{iq{\bar L}n}\chi(q)$,
where $\chi(q)$ is a normalized spinor, we find that this spinor
must obey $H\chi=\lambda\chi$, with
\begin{align}
&H=(e^{-iq{\bar L}}{\bf W}^\dagger+e^{iq{\bar L}}{\bf W})/4\equiv
(A+{\bf B}\cdot \sigmav),\label{H}
\end{align}
where $A=\cos(q{\bar L})\Re d-\sin(q{\bar L})\Im d,\ B_x=0,\
B_y=\cos(q{\bar L})\Im b_y+\sin(q{\bar L})\Re b_y$, and
$B_z=\sin(q{\bar L})\Im b_z-\cos(q{\bar L})\Re b_z$. Therefore,
$\chi$ is an eigenstate of ${\bf n}\cdot\sigmav$, with the unit
vector ${\bf n}={\bf B}/|{\bf B}|$: ${\bf
n}\cdot\sigmav\chi^{}_\mu=\mu\chi^{}_\mu,\ \mu=\pm 1$, and we have
$\lambda=A+\mu |{\bf B}|$. Thus, $\epsilon$ and $q$ must obey the
equation $(\lambda-A)^2=B_y^2+B_z^2$. At fixed $q$, this is a
polynomial of degree 6 in $\epsilon$, so that one expects 6 energy
bands. In the special symmetric case where
$\epsilon^{}_b=\epsilon^{}_c$, partially treated in Ref.
\cite{PRB}, two of these solutions represent dispersionless
solutions at $\epsilon=\epsilon_b$, and thus one is left with only
four bands (whose shape and location varies with $\epsilon_b$).
The value of $\epsilon^{}_a$ only represents a shift in energy, so
we fix $\epsilon^{}_a=0$. For $\epsilon^{}_b>\epsilon^{}_c=0$ the
dispersionless modes become dispersive, and in general the
spectrum also becomes asymmetric with respect to $q
\leftrightarrow -q$ and to $\epsilon \leftrightarrow -\epsilon$.
An example of this phenomenon is shown in Fig. \ref{spec}, for the
square diamond ($\beta=\pi/4$).

As explained in \cite{PRB}, the ballistic conductance at a given
energy $\epsilon$ is equal to $G=(e^2/h)g$, where $g$ is the
number of propagating wave solutions which move in one direction.
To study $g$, we now fix $\epsilon$ and solve the spectrum
equation for $q$. In the general case, this equation turns into a
quartic equation in $\cos(q{\bar L})$. This equation reduces to a
quadratic equation for $\epsilon^{}_b=\epsilon^{}_c$. Out of the
solutions for $q$ we count only the right-moving propagating
modes, with real $q$ and with a positive velocity
$v=\partial\epsilon/\partial q$. As in \cite{PRB}, we again find
ranges of $\epsilon$ with $g=0,\ 1$ or $2$. Generally $g$ depends
on all three parameters $\epsilon,\ \alpha$ and $\phi$. Unlike
Ref. \cite{PRB}, where we fixed $\epsilon$ and presented results
as functions of $\alpha$ or $\phi$, here we fix $\phi$ and Fig.
\ref{contour} presents contour plots of $g(\epsilon,\alpha/\pi)$
at $\phi=\pi/2$. As already indicated by Fig. \ref{spec}, changing
$\epsilon^{}_b$ can open the large gap which existed in the
symmetric case near $\epsilon=0$. This can be seen by comparing
the top two plots in Fig. \ref{contour}: the right hand side plot,
for $\epsilon^{}_b=J$ and $\epsilon^{}_c=0$, is asymmetric with
respect to changing the sign of $\epsilon$, and it exhibits
non-zero ballistic conductance at small positive energy. The gate
voltage governing $\epsilon^{}_b$ can thus be used efficiently to
vary the ballistic conductance between zero and non-zero values.

\section{Spin Filter}

In the regions with $g=1$ we have only one propagating mode. As
explained above, each mode is associated with a single spinor
$\chi(q)$. The spinor of an electron coming from the left will in
principle be written as a linear combination of all the four (or
six) solutions. However, when $g=1$ all the right moving modes
except one are evanescent, i.e. they decay with distance.
Therefore, the regions with $g=1$ represent full polarization of
the conducting electrons. For the symmetric case, Fig.
\ref{contour} also compares between three values of the rhombus
angle $\beta$. Interestingly, the square diamond ($\beta=\pi/4$)
is not the optimal filter; increasing $\beta$ broadens the regions
with $g=1$, where we have full polarization of the electrons.
Therefore, we present below results for $\beta=.35\pi$.


\begin{figure}[h]
\begin{center}
\end{center}
\caption{(Supplied separately) Contour plots in the
$\epsilon-\alpha/\pi$ plane of the ballistic conductance $g$. (a)
$\beta=\pi/4$ and $\epsilon^{}_b=0$, (b) $\beta=\pi/4$ and $\epsilon^{}_b=J$. (c) $\epsilon^{}_b=0$ and $\beta=.15\pi$. (d)
$\epsilon^{}_b=0$ and $\beta=.35\pi$. All other site energies are zero, and all
plots have $\phi=\pi/2$. The values $g=0,\ 1$ and $2$ are
represented by dark, medium and bright areas.} \label{contour}
\end{figure}

\begin{figure}[h]
\begin{center}
\includegraphics[width=3.6cm]{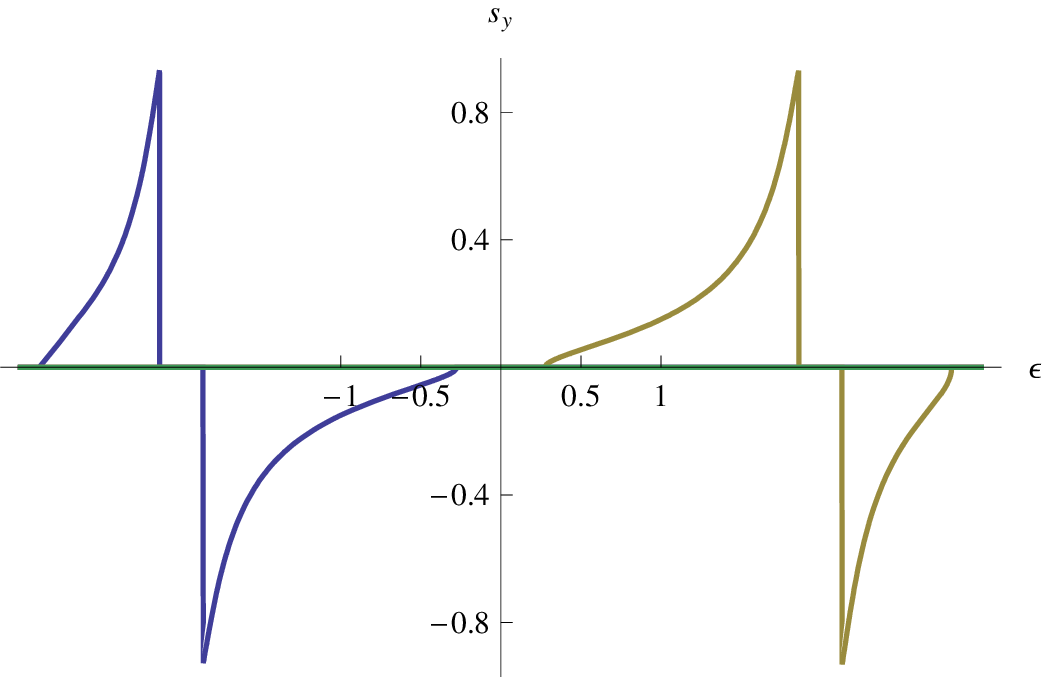}
\hspace{0.3cm}\includegraphics[width=3.6cm]{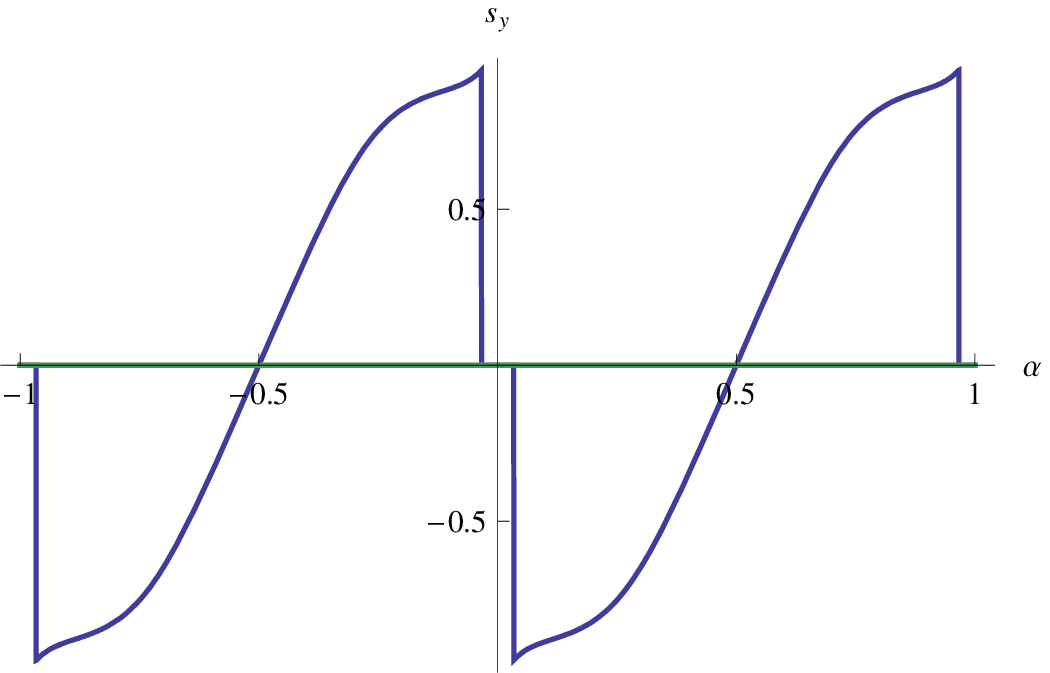}\\
\vspace{0.4cm}\includegraphics[width=3.6cm]{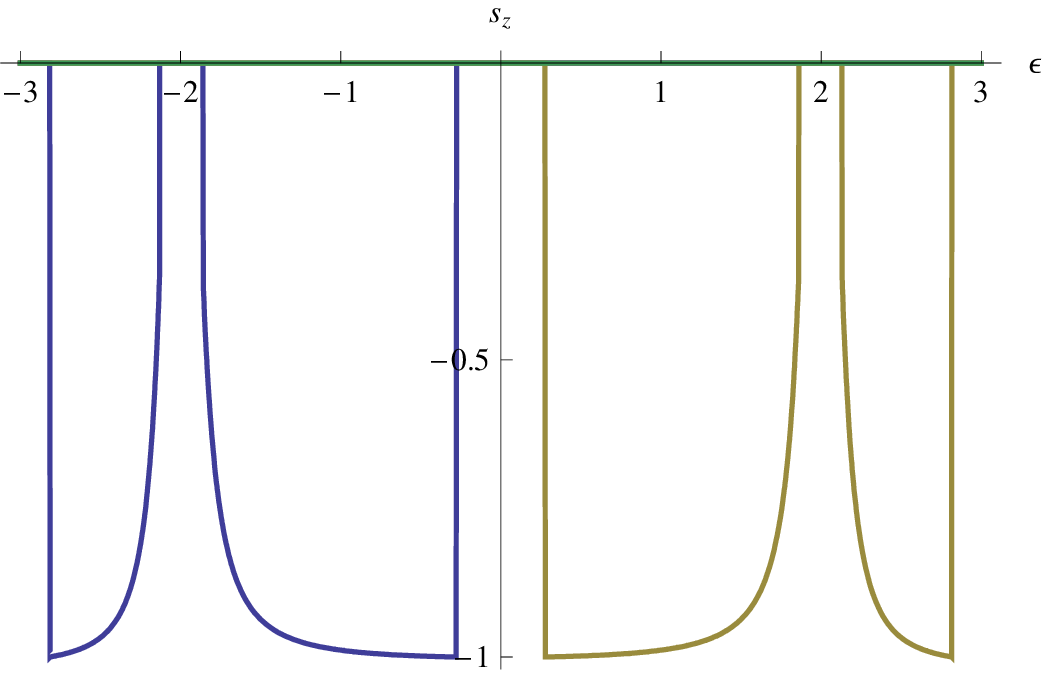}
\hspace{0.3cm}\includegraphics[width=3.6cm]{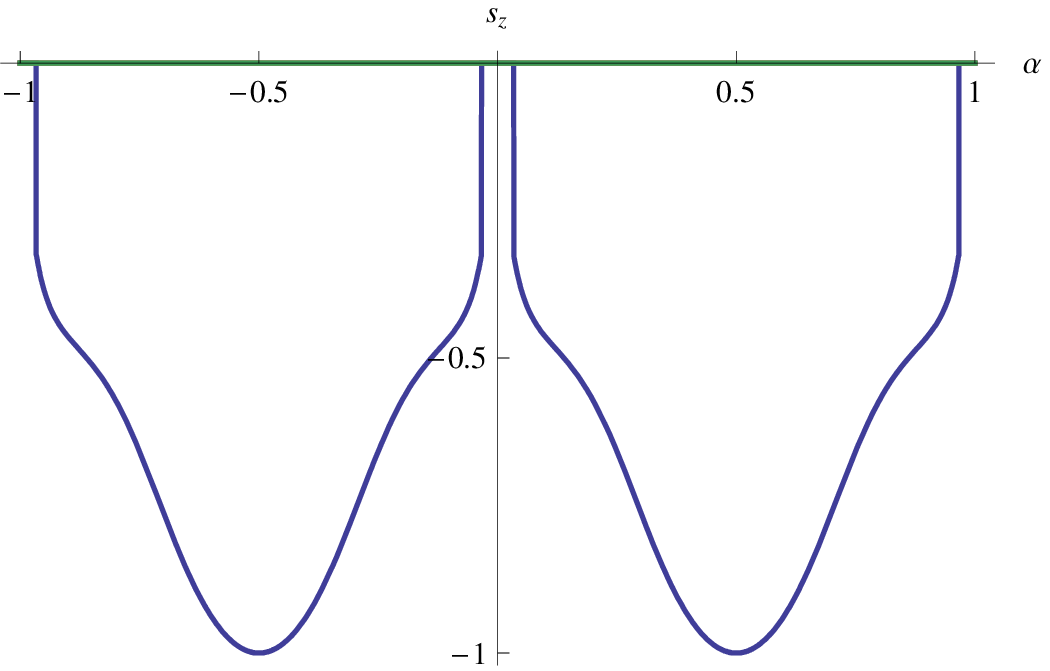}\\
\vspace{.4cm}\includegraphics[width=3.6cm]{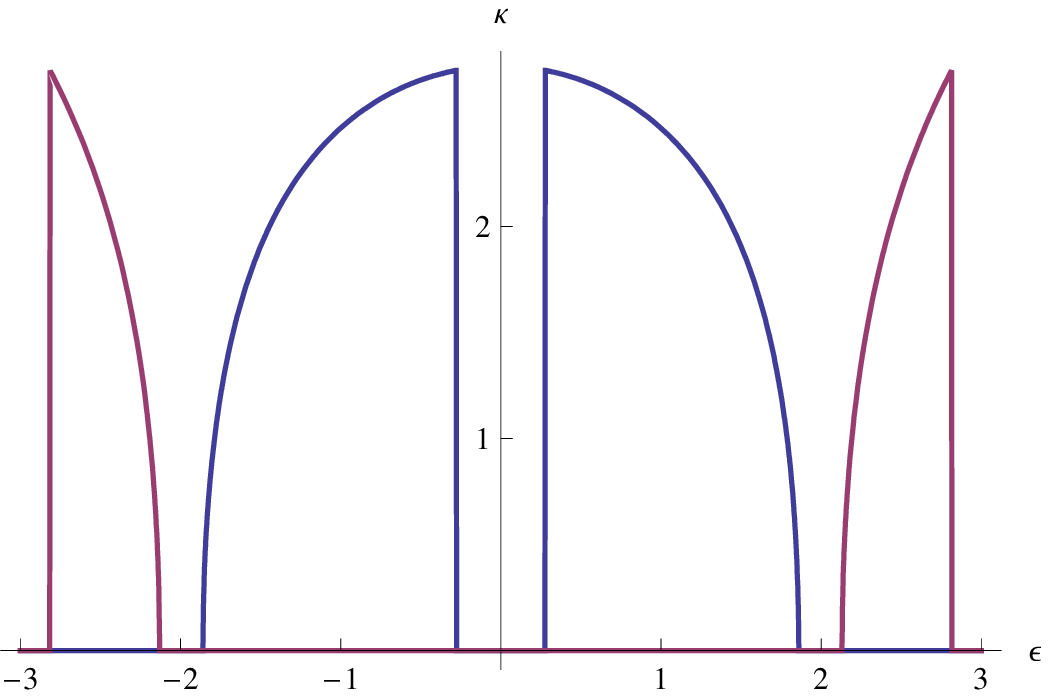}
\hspace{0.3cm}\includegraphics[width=3.6cm]{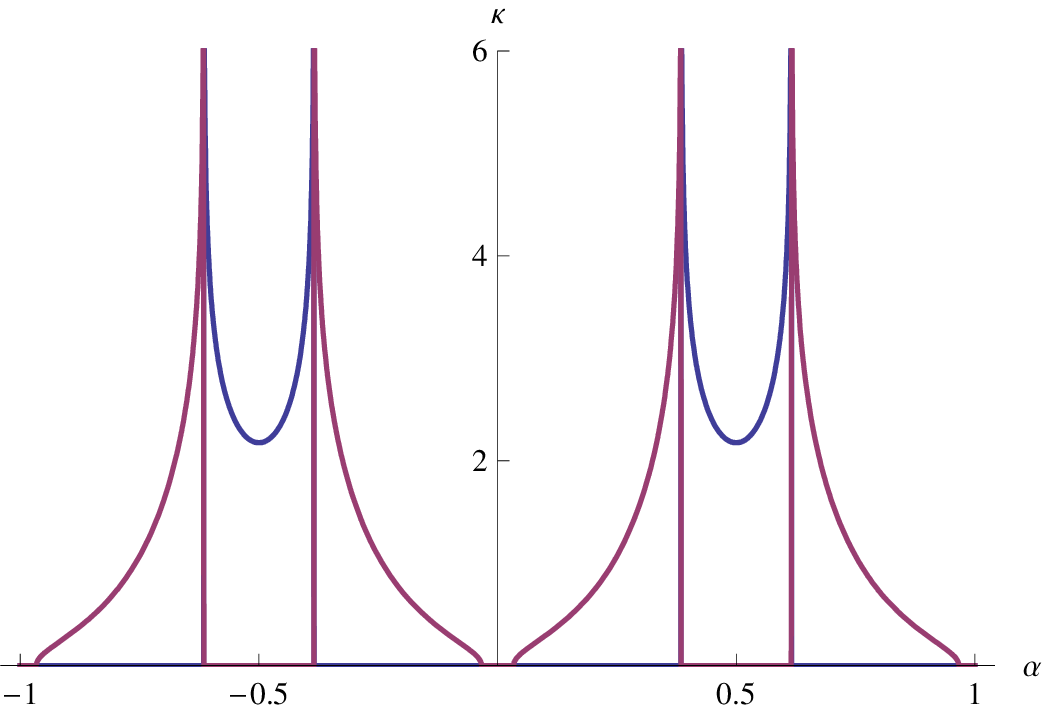}\\
\end{center}
\caption{Spin components $S_y$ and $S_z$ of the polarized
propagating mode and the evanescent inverse decay length (in units
of $1/{\bar L}$) versus $\epsilon$ at $\alpha=.45\pi$ (left) and
versus $\alpha/\pi$ at $\epsilon=-1.1J$ (right). All plots have
$\beta=.35\pi$. Data are shown only in regions where $g=1$.}
\label{kapspin}
\end{figure}

Looking at each panel in Fig. \ref{contour}, we can identify cuts
for which there are broad regions with $g=1$. For each cut, the
spin polarization is given by $\langle
\chi|\sigmav|\chi\rangle=\mu {\bf n}$. Figure \ref{kapspin} shows
these spin components, as a function of $\epsilon$ at fixed
$\alpha=.45\pi$ (the spins are fully aligned in the $z-$direction
for $\alpha=.5\pi$) and as a function of $\alpha$ at fixed
$\epsilon=-1.1J$. As one can see, small changes in $\epsilon$
(determined by the Fermi energy) or in $\alpha$ (determined by the
voltage which fixes the strength of the Rashba SOI) can cause
jumps in $S_y$ between large positive and negative values.

In practical situations, one will not use an infinite chain of
diamonds. As discussed in Ref. \cite{PRB}, the spin polarization
is maintained even for a finite chain, provided we use adiabatic
contacts at the exit. However, this finite chain must be long
enough so that the evanescent modes will decay before the
electrons come out. In the symmetric case, $\cos(q{\bar L})$ is
found from solving the quadratic equation $(\lambda-A)^2=|{\bf
B}|^2$. For $g=1$, one of the solutions has $|\cos(q{\bar L})|\le
1$, and therefore a real $q$, while the other solution has
$|\cos(q{\bar L})|> 1$, and therefore an imaginary $q=i\kappa$.
The bottom panels in Fig. \ref{kapspin} show the imaginary part of
$q$ for these other modes, denoted by $\kappa$. As one can see,
there are broad regions in which $\kappa>1$, so that a small
number of diamonds suffices for the evanescent modes to decay.
Interestingly, $\kappa$ diverges to infinity as $\alpha$
approaches special values, for which the coefficient of
$\cos^2(q{\bar L})$ in the above quadratic equation approaches
zero. Writing this equation as $ax^2+bx+c=0$, the solutions for
small $a$ are $\cos(q{\bar L})=x_1\approx -b/c$ and
$\cosh(\kappa{\bar L})=x_2\approx -b/a$. Clearly, $\kappa$
diverges as $a$ approaches zero. Using the specific relations
which follow Eq. (\ref{H}), we identify $a=d^2+b_y^2-b_z^2$. It is
then easy to check that the condition $a=0$ is equivalent to
$\det{\bf W}=\det{\bf W}^\dagger=0$. When this condition is
obeyed, ${\bf W}$ and ${\bf W}^\dagger$ have a vanishing
eigenvalue, which means that the corresponding eigenvector is
completely blocked [see Eq. (\ref{ren})]. Interestingly, Ref.
\cite{molnar2} found similar blocking for finite chains at special
values of the parameters.

\section{Tight-binding versus continuum models}

\subsection{Tight-binding model with many sites}

We now generalize our model, by replacing each edge of a diamond
by $M$
bonds in series. 
Consider a single `large' bond, e.g. the bond $ab$ in Fig.
\ref{1}. Solving the appropriate one-dimensional tight-binding
model along this bond, we find that the spinor on the site $m$,
$\psi^{(ab)}(m)$ [$0\le m\le M$, with
$\psi^{(ab)}(m=0)\equiv\psi^{}_a$ and
$\psi^{(ab)}(M)\equiv\psi^{}_b$], is given by
\begin{align}
\psi^{(ab)}(m)&=\frac{\sin[ka(M-m)]}{\sin(kaM)}[U^{\dagger}_{ab}]^{m/M}\psi^{}_a\nonumber\\
&+\frac{\sin(kam)}{\sin(kaM)}[U^{}_{ab}]^{(M-m)/M}\psi^{}_b,\label{psiab}
\end{align}
where $U^{}_{ab}$ is the same as in Eq. (\ref{eq1}), $a=L/M$ is
the new lattice constant, and $k$ is the wave vector for the
one-dimensional solution, related to the energy $\epsilon$ and to
the new elementary hopping energy $J$ via $\epsilon=-2J\cos(ka)$.

We next discuss a general node on the network, $u$. The
tight-binding equation at this node is
\begin{align}
(\epsilon-\epsilon^{}_u)\psi^{}_u=-J\sum_v
[U_{uv}]^{1/M}\psi^{(uv)}(1),\label{eq7}
\end{align}
where we assumed that all the `large' bonds are equivalent to each
other, having the same number $M$ of internal bonds, the same
lattice constant $a$ and the same hopping energy $J$. Substituting
Eq. (\ref{psiab}) on the right hand side, it is straightforward to
rewrite Eq. (\ref{eq7}) as
\begin{align}
E_u\psi^{}_u=-J\sum_vU^{}_{uv}\psi^{}_v,\label{neweq}
\end{align}
with \cite{JPC}
\begin{align}
E_u=&-J\Bigl
(N_u\cos(kaM)-(N_u-2)\sin(kaM)\cot(ka)\nonumber\\
&+\epsilon^{}_u\sin(kaM)/[J\sin(ka)]\Bigr ),\label{Eu}
\end{align}
where $N_u$ is the number of bonds meeting at site $u$. The
generalization to arbitrary bonds is obvious. Equations
(\ref{neweq}) look exactly like our tight-binding equations
(\ref{eq2}), provided we replace $\epsilon-\epsilon^{}_u$ by
$E_u$. Therefore, we might expect some similarities in the
solutions.

The equation for $\psi^{}_b(n)$ in (\ref{eq2}) now has $E_b$ on
the left hand side, and therefore the dispersionless modes contain
all the solutions of $E_b=0$. With $\epsilon^{}_b=0$, this yields
$\epsilon=-2J\cos(ka),\ ka=(n+1/2)\pi/M$. Except for these
energies, we again eliminate the side site spinors, restricting
ourselves to the symmetric case, $\epsilon^{}_b=\epsilon^{}_c$. In
this case, we also have $\gamma^{}_b=\gamma^{}_c\equiv\gamma$. It
is then convenient to separate the common factor $\gamma$ from Eq.
(\ref{db}), and rewrite the generalized Eq. (\ref{ren}) in the
form
\begin{align}
4\Lambda\psi^{}_a(n)={\tilde{\bf
W}}^\dagger\psi^{}_a(n-1)+{\tilde{\bf W}}\psi^{}_a(n+1),
\end{align}
with  ${\tilde{\bf W}}\equiv{\bf W}/\gamma$, and with
$4\Lambda=E_a/\gamma-4$, where now $\gamma=J^2/E_b$. The symmetric
cases which we described above are characterized by the same
matrix ${\tilde{\bf W}}$.   
Since all the spin physics described above resulted only from the
matrix ${\tilde{\bf W}}$, which does not depend on the energy
$\epsilon$, all of that discussion will remain unchanged. The only
effect of adding the internal bonds on each `large' bond appears
in the new parameter $\Lambda$, which is now given by
\begin{align}
\Lambda=E_aE_b/(4J^2)-1.\label{Lam}
\end{align}
As before, the spectrum is determined by \begin{align}
\Lambda={\tilde A}\pm|{\tilde{\bf B}}|,\label{newspec}
\end{align}
 where
${\tilde A}=A/\gamma$ and ${\tilde{\bf B}}={\bf B}/\gamma$. For
each value of $q$, Eq. (\ref{newspec}) determines two values for
$\Lambda$, which are independent of $M$ and of $\epsilon$. We then
solve each of these equations for all possible values of $ka$, and
obtain the energies of the various bands via
$\epsilon=-2J\cos(ka)$.

\begin{figure}[h]
\begin{center}
\includegraphics[width=3.6cm]{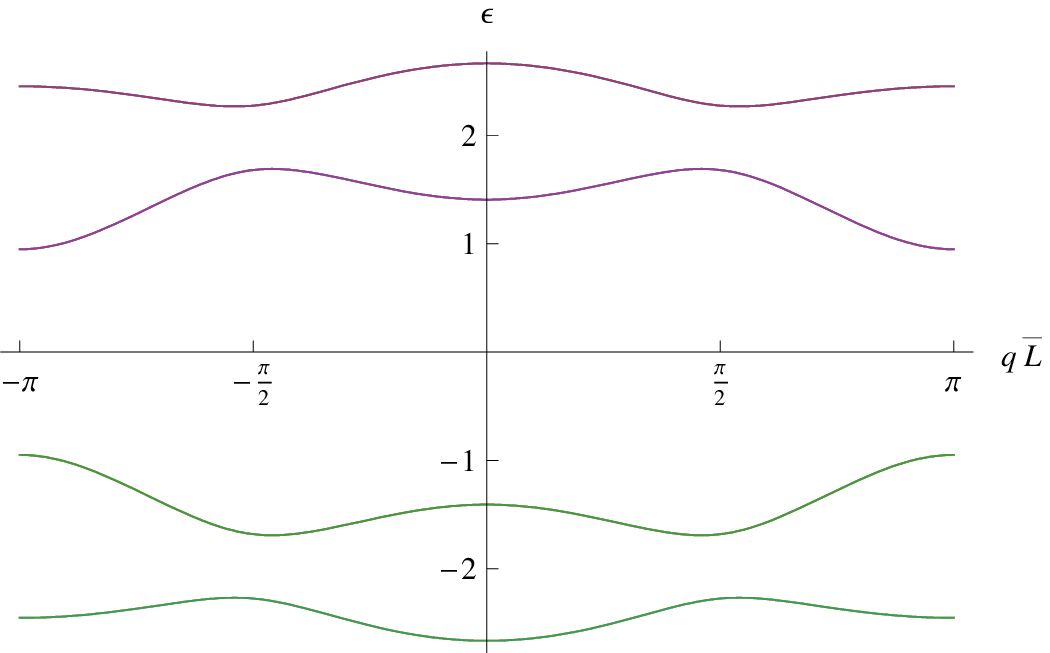}\ \ \includegraphics[width=3.6cm]{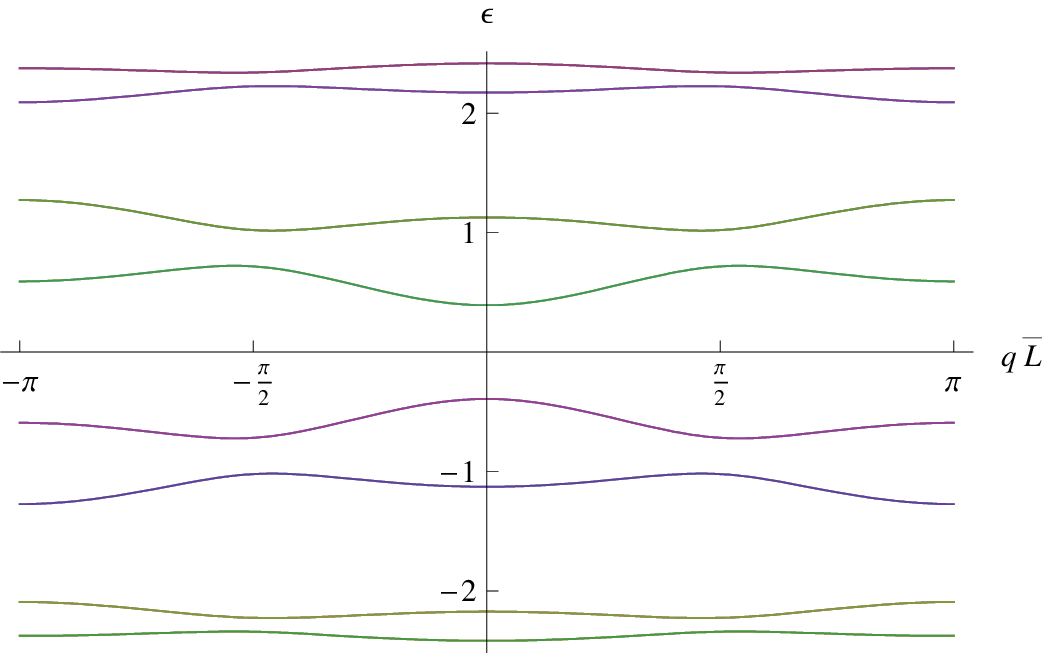}\\
\vspace{0.3cm}
\includegraphics[width=3.6cm]{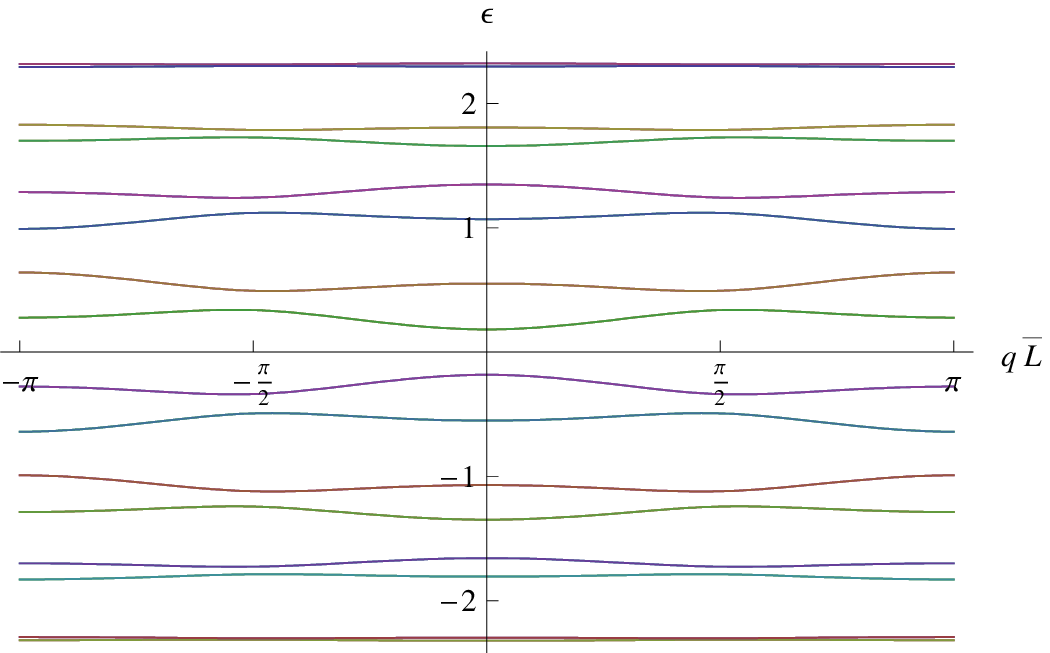}\ \
\includegraphics[width=3.6cm]{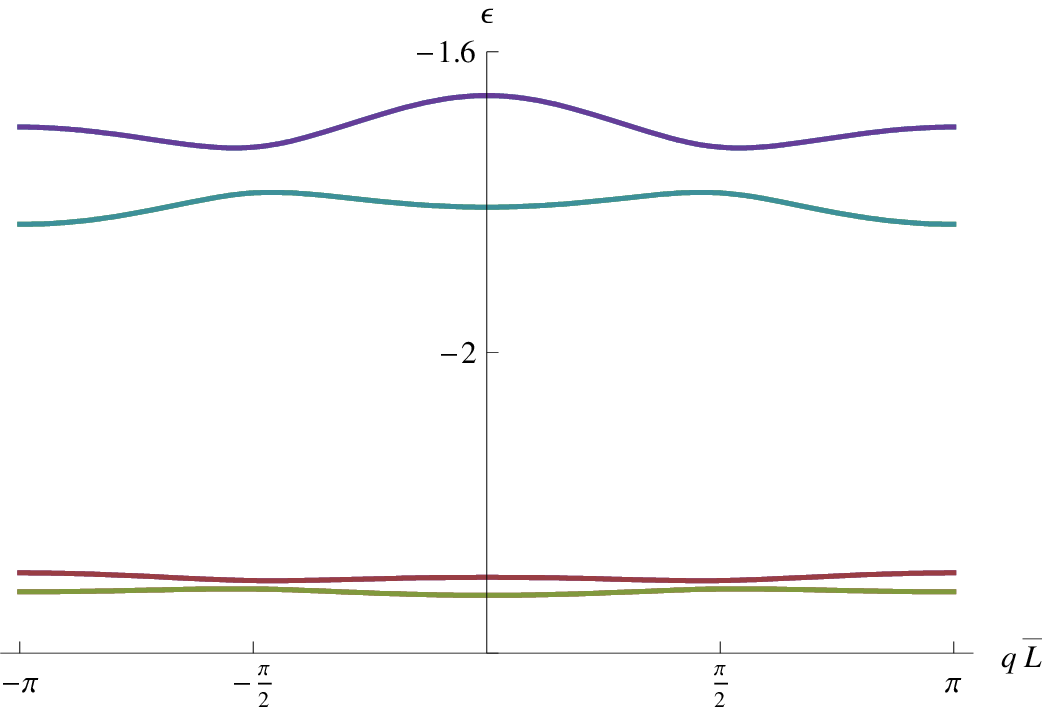}
\end{center}
\caption{Spectra for the diamond chain with $M$ bonds on each
`large' bond, with $\beta=\pi/4,\ \phi=\pi/2,\ \alpha=.4\pi$. Top:
$M=1$ (left), $M=2$ (right). Bottom: $M=4$. All energies are in
units of $J$.} \label{newE}
\end{figure}

For the special case
$\epsilon^{}_a=\epsilon^{}_b=\epsilon^{}_c=0$, Eq. (\ref{Lam})
reduces to $\Lambda=\cos(2kaM)-.5\sin(2kaM)\cot(ka)$. Examples of
the spectra for this case and for several values of $M$ are shown
in Fig. \ref{newE}. Clearly, the number of bands increases with
$M$. However, the basic qualitative shapes of $\epsilon(q)$ for
all the bands are similar to each other. Thus, we expect that
within each band we would reproduce the filtering properties
described in Sec. 3. However, the energy scales for each band
become narrower as $M$ increases. This narrowing is most
pronounced for the lowest band, which always appears below the
band of the one-dimensional solutions, $\epsilon<-2J$. Within our
tight-binding model, this narrow band involves an imaginary value
of $k$, $\kappa=ik$, implying very small wave functions in the
middle of each `large' bond. This imaginary solution results from
the solution of $\Lambda=\cosh(2\kappa aM)-.5 \sinh(2\kappa
aM)\coth(\kappa a)$. At large $M$, $\Lambda$ becomes negligible,
and the two bands converge to a single band, with $\coth(\kappa a)
\rightarrow 2$ and thus $\epsilon=-2J\cosh(\kappa a) \approx
-2.3094 J$. The bottom panels in Fig. \ref{newE} show the spectrum
for $M=4$. The right hand side panel zooms on the two lowest
bands, demonstrating that these bands maintain their qualitative
features even as $M$ increases. The lowest band does indeed narrow
down, becoming dispersionless in the limit $M\rightarrow \infty$,
where the equation $\Lambda=0$ yields the solution
$\epsilon=-2\cosh(\kappa a)$, with $\coth(\kappa a)=2$.

\subsection{Continuum model}

\begin{figure}[h]
\begin{center}
\includegraphics[width=3.6cm]{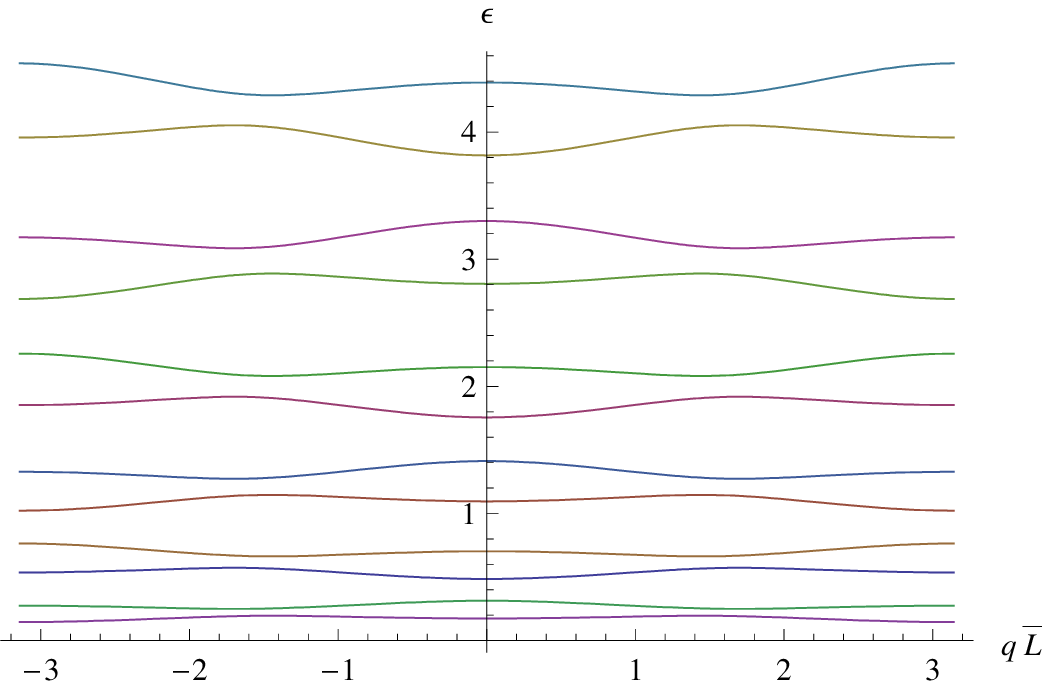}\ \ \includegraphics[width=3.6cm]{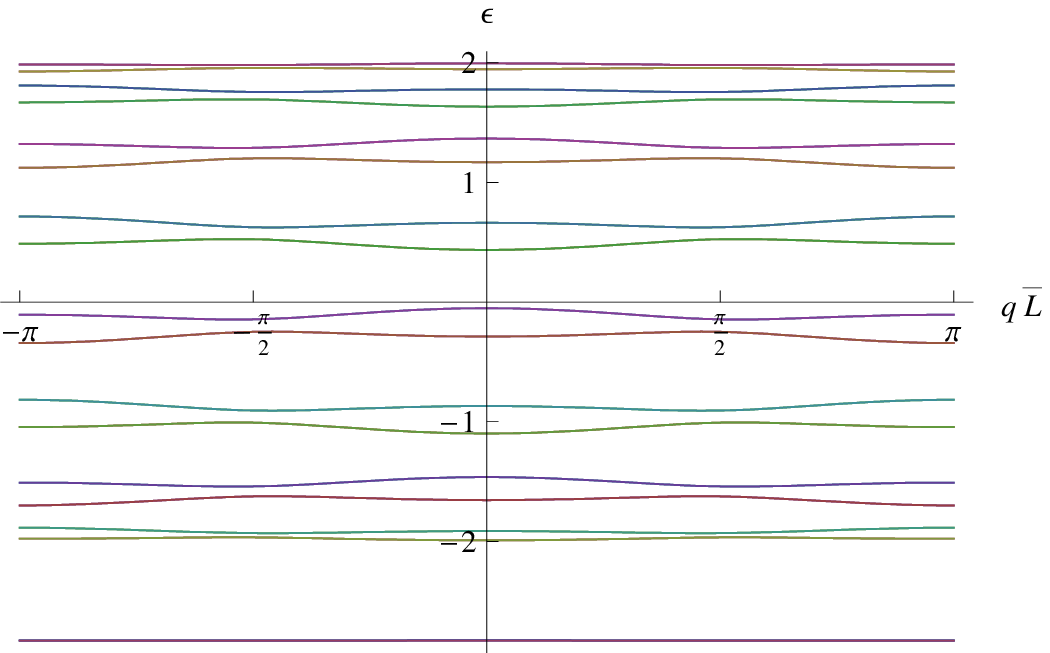}\\
\end{center}
\caption{Spectra for the diamond chain for the continuum model
(left, with arbitrary energy units) and for the modified
tight-binding ($M=4,\ \epsilon^{}_a=2J$, right), with
$\beta=\pi/4,\ \phi=\pi/2,\ \alpha=.4\pi$ and with energies in
units of $J$.} \label{BERC}
\end{figure}

In contrast to our calculations, Refs. \cite{berc1} and
\cite{berc2} used a continuum wire model for each `large' bond on
each diamond. Their solution for the spinor at distance $x$ from
node $a$ on the bond $ab$ is given by
\begin{align}
\psi^{(ab)}(x)&=\frac{\sin[k(L-x)]}{\sin(kL)}[U^{\dagger}_{ab}]^{x/L}\psi^{}_a\nonumber\\
&+\frac{\sin(kx)}{\sin(kL)}[U^{}_{ab}]^{(L-x)/L}\psi^{}_b.\label{psiabB}
\end{align}
Since the electron on each bond is now free, its energy is given
by $\epsilon=\hbar^2k^2/(2m^\ast)$, and Bercioux {\it et al.} plot
$\epsilon=k^2$. As usual with tight-binding equations, this
solution coincides with our tight-binding solution (\ref{psiab})
in the limit $M\rightarrow\infty$, keeping $a=L/M\rightarrow 0$
and $x=ma$. Having found these solutions, Refs. \cite{berc1} and
\cite{berc2} proceed to use the Neumann boundary conditions at the
nodes:
\begin{align}
\sum_v\frac{\partial\psi^{(uv)}(x)}{\partial x}\Big |_{x=0}=0.
\end{align}
With these conditions, they end up with equations like our
(\ref{neweq}), but with our $E_u$ replaced by $-JN_u\cos(kL)$.
Thus, they would have $\Lambda=\cos(2kL)$, and therefore
$k=(n\pi\pm.5\arccos\Lambda)/L$. These results, with
$\epsilon=k^2/100$, are reproduced in the left panel of Fig.
\ref{BERC}. Again, the qualitative shape of each band looks
similar to ours. However, this approach cannot reproduce the
lowest band with imaginary $k$ which we found for finite $M$.

As discuss in Ref. \cite{JPC}, there is no way to make the
tight-binding model and the continuum models identical for any
finite $M$. However, in the limit $M\rightarrow\infty$ one can
 modify the tight-binding model,
by setting a site energy $\epsilon^{}_a=2J$. With this special
value, our Eq. (\ref{Lam}) becomes
$\Lambda=\cos(2kL)+\sin(2kL)\sin(ka)$, which becomes identical to
the expression used in Refs. \cite{berc1} and \cite{berc2} when
$a\rightarrow 0$. Unfortunately, we know of no good reason to
choose this particular value for the site energy in the
tight-binding model. The right hand panel in Fig. \ref{BERC} shows
results of the modified tight-binding model for $M=4$. Except for
the dispersionless band below the continuum, we expect these
results to approach those in the left hand side panel as
$M\rightarrow\infty$.

\section{Discussion}

Reference \cite{PRB} already compared the results of a chain of
diamonds to a single diamond. As also noted in Ref.
\cite{molnar2}, having more loops in series broadens the parameter
regions which yield full polarization. Reference \cite{PRB} also
discussed the conditions for having full filtering on a chain of
finite length. Here we have extended that discussion by showing
that there exist broad regions with relatively short evanescent
decay lengths, so that one can obtain filtering with relatively
short chains.

In addition, we have shown that the filtering results are quite
robust: in addition to the parameters discussed in Ref.
\cite{PRB}, the filtering persists upon changing many additional
parameters (e.g. the opening angle of each diamond). We also
mention the asymmetry of the spectra in the non-symmetric case,
which implies that a large (positive or negative) bias voltage
between the left and right hand ends of the device can yield
different currents in the two directions.

We hope that the present discussion will stimulate attempts to
realize our filter experimentally.

 \vspace{.5cm}

\noindent{\bf Acknowledgements.} We acknowledge discussions with
Joe Imry. AA and OEW acknowledge the hospitality of NTT and of the
ISSP, where this project started, and support from the ISF and
from the DIP.

\end{document}